\documentclass[conference]{IEEEtran}
\usepackage[utf8]{inputenc}
\usepackage{subcaption}
\usepackage{caption}
\usepackage{amsmath}
\usepackage{amsfonts}
\usepackage{braket}
\usepackage{mathtools}
\usepackage{graphicx}
\usepackage{xcolor}
\usepackage{url}

\newcommand\blfootnote[1]{
    \begingroup
    \renewcommand\thefootnote{}\footnote{#1}
    \addtocounter{footnote}{-1}
    \endgroup
}

\ifCLASSINFOpdf
\else
\fi
\begin{document}
\title{Quantum Resources for Pure Thermal Shadows}

\author{\IEEEauthorblockN{Arnav Sharma}
\IEEEauthorblockA{
\textit{Lincoln Laboratory} \\
\textit{Massachusetts Institute of Technology}\\
Lexington, MA USA \\
Arnav.Sharma@ll.mit.edu
}
\and
\IEEEauthorblockN{Kevin M. Obenland}
\IEEEauthorblockA{
\textit{Lincoln Laboratory} \\
\textit{Massachusetts Institute of Technology}\\
Lexington, MA USA \\
Kevin.Obenland@ll.mit.edu
}}

\maketitle

\begin{abstract}
Calculating the properties of Gibbs states is an important task in Quantum Chemistry and Quantum Machine Learning. Previous work has proposed a quantum algorithm which predicts Gibbs state expectation values for $M$ observables from only $\log{M}$ measurements, by combining classical shadows and quantum signal processing for a new estimator called \textit{Pure Thermal Shadows}. In this work, we perform resource analysis for the circuits used in this algorithm, finding that quantum signal processing contributes most significantly to gate count and depth as system size increases. The implementation we use for this also features an improvement to the algorithm in the form of more efficient random unitary generation steps. Moreover, given the ramifications of the resource analysis, we argue that its potential utility could be constrained to Fault Tolerant devices sampling from the Gibbs state of a large, cool system.
\end{abstract}

\blfootnote{DISTRIBUTION STATEMENT A. Approved for public release. Distribution is unlimited.
This material is based upon work supported by the Under Secretary of Defense for Research and Engineering under Air Force Contract No. FA8702-15-D-0001. Any opinions, findings, conclusions or recommendations expressed in this material are those of the author(s) and do not necessarily reflect the views of the Under Secretary of Defense for Research and Engineering.
© 2024 Massachusetts Institute of Technology.
Delivered to the U.S. Government with Unlimited Rights, as defined in DFARS Part 252.227-7013 or 7014 (Feb 2014). Notwithstanding any copyright notice, U.S. Government rights in this work are defined by DFARS 252.227-7013 or DFARS 252.227-7014 as detailed above. Use of this work other than as specifically authorized by the U.S. Government may violate any copyrights that exist in this work.}

\section{Introduction}
Estimating properties of an unknown quantum state is an important and nontrivial task, both for experimental purposes and quantum algorithms. Gibbs state properties are particularly useful for several different applications ranging from materials science to optimization and machine learning. Gibbs states are mixed quantum states which describe systems in thermodynamic equilibrium with their environment at finite temperatures; we can define a Gibbs state for a given system using a density matrix of the form:

\begin{equation}
\rho_\beta = e^{-\beta H}/Z
\end{equation}
where $ Z = \text{Tr}e^{-\beta H}$ is the partition function, $\beta$ is inverse temperature, and $H$ is the system Hamiltonian. Explicitly calculating these states and their properties is difficult to do classically due to the exponentially growing Hilbert space along with a sign
problem that makes simulating fermionic systems difficult~\cite{loh}. Some  classical algorithms have developed alternative techniques, 
including Monte Carlo methods which avoid the sign problem and sampling methods whose runtimes scale polynomially with system size at high temperatures~\cite{yin}. A number of quantum approaches to Gibbs state preparation and sampling algorithms exist as well, including a quantum Metropolis algorithm which features a quadratic speedup over its classical counterpart~\cite{yung}, as well as an algorithm which scales with the square root of Hilbert Space size and inverse temperature~\cite{chowdhury}. However, these algorithms require large, Fault Tolerant devices which do not yet exist. Considering near-term NISQ (Noisy Intermediate-Scale Quantum) devices, variational methods for Gibbs state preparation have been proposed and implemented experimentally~\cite{consiglio}. These methods have the advantage of only requiring a polynomially number of operations, however, they are also susceptible to the barren plateau problem.

The approach of~\cite{Coopmans} proposes to apply the classical shadow protocol to thermal pure quantum (TPQ) states, which themselves are prepared through Quantum Signal Processing, as a way to estimate Gibbs state properties. These particular shadows were referred to as \textit{Pure Thermal Shadows}, and their ability to approximate Gibbs state expectation values as well as train a Quantum Boltzmann Machine were numerically verified. In this work, we aim to validate this algorithm, analyze it for potential improvements, and evaluate how its resource requirements might scale for use on an actual quantum device.

\section{Background}
\label{section:bg}
\subsection{Classical Shadows}
\label{section:CS}
 To make reasonable estimations of an unknown state's properties, traditional methods like quantum state tomography require an exponential number of measurements with respect to system size. Shadow tomography seeks to describe not an entire quantum state, but a "shadow" of it on a set of measurements~\cite{Aaronson}. Leveraging this,~\cite{Huang2020} proposed a $\textit{classical}$ description of quantum states called $\textit{classical shadows}$, which can accurately predict $M$ different functions (expectation values for particular observables) of the quantum state with only $O(\log{M})$ measurements.
 
Classical shadows can be constructed from randomized measurements of a quantum state, and used to efficiently predict many of its properties. Consider an $n$-qubit state $\rho$, prepared by some circuit (in this case our Gibbs state). The procedure works by applying some unitary transformation $V$, chosen randomly from some ensemble, to $\rho$ before measuring the resulting state ($V\rho V^{\dagger}$) in the computational basis giving an outcome $\ket{b}$. 

From the measurement outcome $\ket{b}$, we store the 'reverse' operation of our random measurement in classical memory: $V^{\dagger} \ket{b}\bra{b} V$. We call this value a 'snapshot' of our state. The average of these snapshots defines a quantum measurement channel, $\mathcal{M}$. Thus, we can construct a Classical Shadow of $\rho$ with respect to $V$ and $\ket{b}$ like so:

\begin{equation}
\hat{\eta}_{V,b} = \mathcal{M}^{-1}(V^{\dagger}\ket{b}\bra{b} V)
\end{equation}

The inverse of $\mathcal{M}$ is a linear map, which depends on the ensemble from which $V$ is sampled. (Note, this linear map is not CPTP, but since we only use it for classical post-processing this does not pose a problem.) Two common ensembles used for generating random unitaries are:

\begin{enumerate}
    \item Random Clifford unitaries: $V \in Cl(2^n)$
    \item Random Pauli basis transformations: $V = V_1 \otimes ... \otimes V_n \in Cl(2)^{\otimes n}$.
\end{enumerate}
The first is equivalent to generating a random Clifford circuit (for which a polynomial time algorithm exists) while the second is equivalent to transforming each qubit's state into a randomly chosen Pauli-X, Y, or Z basis; that is, each $V_j$ is randomly sampled from $\{H, HS^{\dagger}, I\}$ where $H$ is the Hadamard gate and $S$ is the phase gate. For Clifford measurements, $\hat{\eta}_{V,b} = (2^n + 1)V^{\dagger}\ket{b}\bra{b} V - I$ while for Pauli basis transformations $\hat{\eta}_{V,b} = \bigotimes^{n}_{j=1}(3 V^{\dagger}_j\ket{b_j}\bra{b_j} V_j - I) $. The benefit to choosing the first ensemble is an increased accuracy for estimations, but this comes at the cost of a polynomially increasing circuit depth with respect to system size, while the second ensemble can be implemented with a constant depth circuit.

The state $\rho$ can be classically reconstructed as the expectation value of our shadows over the random unitaries: $\rho = \mathbb{E}_{V,b}[\hat{\eta}_{V,b}]$. Thus, for any of our observables $O_j$ in a set $O$ :
\begin{multline}
    \braket{O_j}_{\rho} = \text{Tr}(\rho O_j) = \text{Tr}(\mathbb{E}[\hat{\eta}_{V,b}] O_j) \\ = \mathbb{E}[\text{Tr}(\hat{\eta}_{V,b} O_j)] = \mathbb{E}_{V,b}[\braket{O_j}_{\hat{\eta}_{V,b}}]
\end{multline}
In other words, the expectation value of an operator for a given state can be estimated by an average of the expectation values for many classical shadows. 

Now, consider the question of \textit{how many} classical shadows we should use for a given expectation value estimation, and \textit{how} we should average them. The classical shadow protocol typically employs a median of means estimator, where a sample of $n_s$ shadows is broken into $K$ equally sized sets of size $S$ and we take the median of the $K$ set means (each calculated from $S$ shadows). The values of $S$ and $K$ are directly related to the performance of our estimation and our observables. Specifically:
\begin{equation}
S = \frac{34\sigma^2}{\epsilon^2}
\end{equation}
where $\epsilon$ is our desired error bound (maximum difference between the estimated expectation value and true one). Also note that for random Pauli measurements $\sigma^2 \leq 3^{\text{locality}(O)}$.

For $K$ we have:
\begin{equation}
K = (2)\log(2M/\delta) 
\end{equation}
where  $M$ is the number of observables whose expectation values we are estimating, and the probability that the estimations for all observables fall within our error bound is $1-\delta$.

Thus, we can conclude that for a given $\epsilon, \delta$ we need at least
\begin{equation}
\label{shadcomp}
n_s = SK = 68\frac{\sigma^2}{\epsilon^2}\log(2M/\delta)
\end{equation}
shadows. Crucially, this suggests that we can estimate $M$ expectation values with $O(\log M)$ measurements---favorable for many observables. Moreover, the logarithmic dependence on $M$ and $1/\delta$ improves upon an arithmetic mean which requires a linear number of samples with respect to $M$ and $1/\delta$ (see the appendices of~\cite{Huang2020} for proofs of sampling complexities).

\subsection{Pure Thermal Shadows}
\label{section:PTS}
Pure Thermal Shadows are classical shadows of Thermal Pure Quantum (TPQ) states. A TPQ state is randomly sampled and can be utilized for estimating finite sets of observables with respect to a mixed thermal state. As given in~\cite{Sugiura2013}, a TPQ state $\ket{\psi}$ for the thermodynamic Gibbs ensemble obeys:

\begin{equation}
\text{Pr}[\bra{\psi} O_j \ket{\psi} - \text{Tr}\rho_\beta O_j \geq \epsilon] \leq C_\epsilon e^{-\alpha n}
\label{tpqcondition}
\end{equation}
for all $O_j \in O$, where $C_\epsilon$ and $\alpha$ are constants. When $C_\epsilon \approx 4||O_j||/\epsilon^2$ and $e^{-\alpha n} = \text{Tr}\rho_\beta^2$, equation \ref{tpqcondition} is satisfied for all $O_j$ by:

\begin{equation}
\ket{\psi_\beta} = \frac{Ue^{-\beta H/2} \ket{0}}{\bra{0}U^\dagger e^{-\beta H}U\ket{0}}
\label{tpq}
\end{equation}
where $U \in Cl(2^n)$ is randomly sampled from at least a unitary 2-design~\cite{Coopmans}.

Pure Thermal Shadows are thus constructed by taking classical shadows of the state $\ket{\psi_\beta}$ as given by Equation \ref{tpq}. Now, the analysis relating the sample complexity with estimation accuracy (i.e. how many shadows are necessary for a given error threshold) from Section \ref{section:CS} was constructed exclusively with classical shadows in mind. It turns out that a  \textit{tighter} bound actually exists which applies for $\textit{both}$ Classical and Pure Thermal Shadows. Namely:

\begin{equation}
n_s = 27\frac{\sigma^2}{\epsilon^2}\log(M/\delta)
\label{numshads}
\end{equation}
 
where the mean squared error specifically for Pure Thermal Shadows has an added term of $O(e^{-n}$).

\section{Methods and Implementation}
\begin{figure}
    \centering
    \includegraphics[width=8cm]{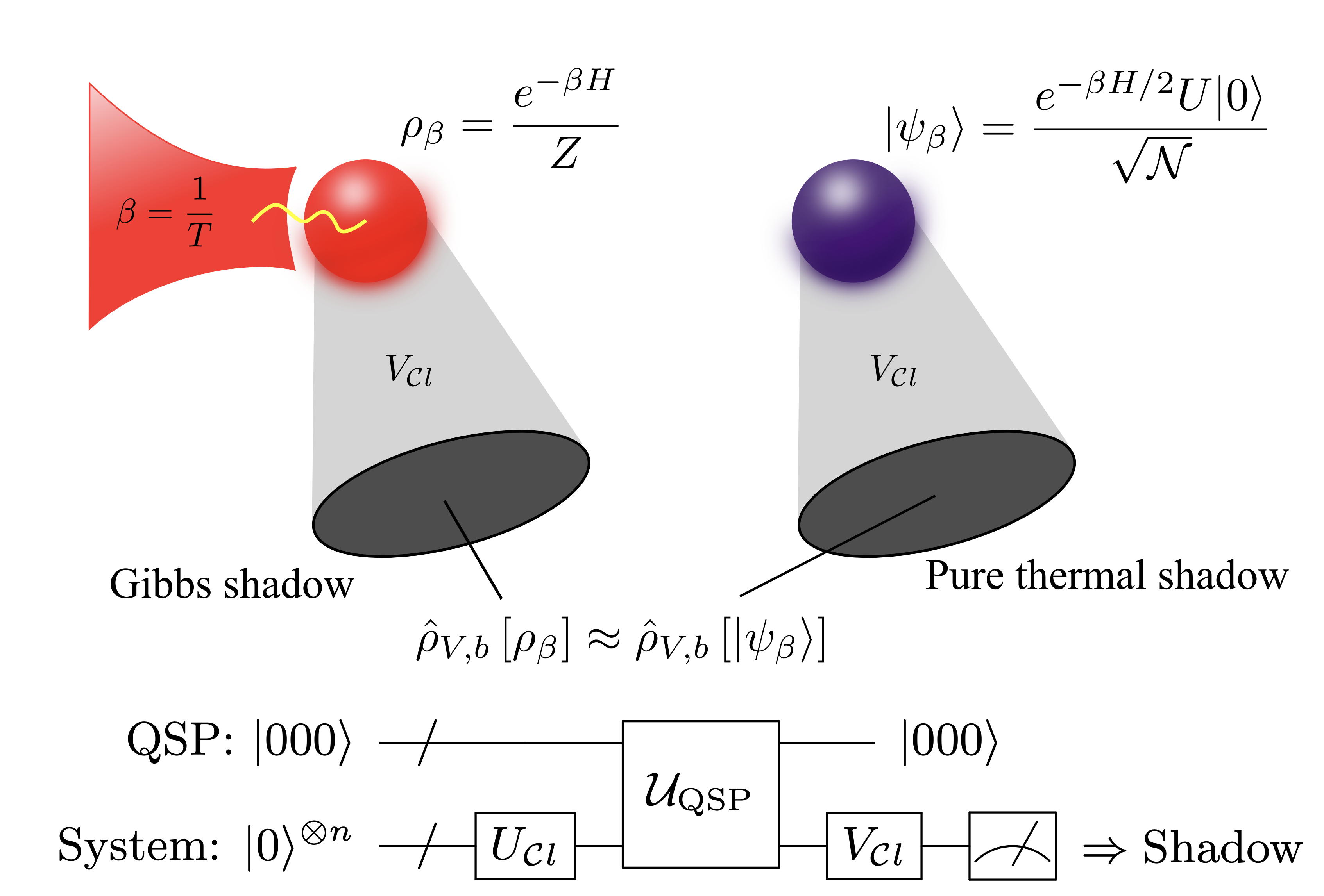}
    \caption{The circuit which generates a Pure Thermal Shadow as outlined in~\cite{Coopmans}. A Thermal Pure Quantum State (Equation \ref{tpq}) is prepared by applying a random unitary ($U_{Cl}$ in the figure) to system qubits $\ket{0}^{\otimes n}$, then performing Quantum Signal Processing (requiring some ancillae shown in the top register) to approximately apply the non-unitary operator $e^{-\beta H /2}$. Finally, the system qubits are measured in a random basis (transformed via $V_{Cl}$) to produce a Pure Thermal Shadow. These shadows can approximate the classical shadows of a Gibbs state insofar as estimating expectation values for a set of observables.}
    \label{fig:PTS}
\end{figure}

\begin{figure*}
    \includegraphics[width=17cm]{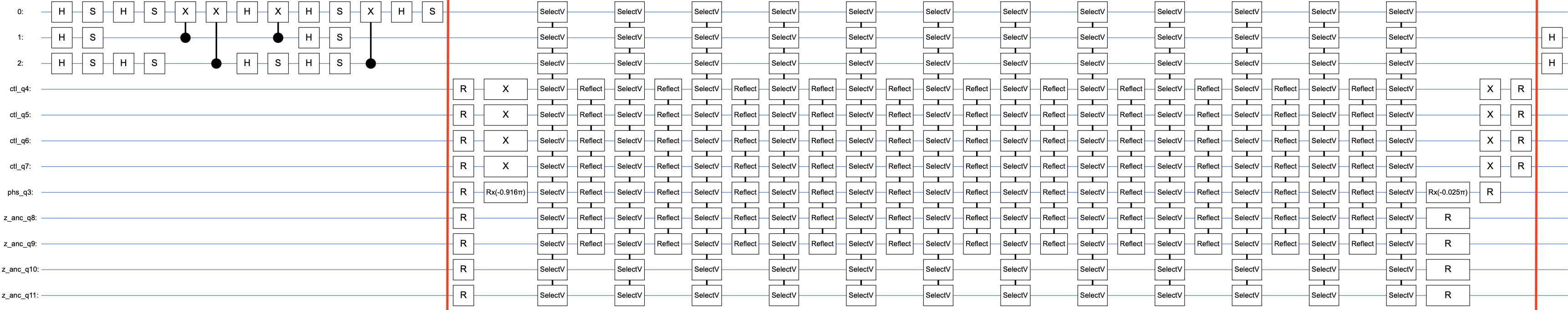}
    \caption{Example quantum circuit constructed in Cirq representing the measurement of a single Pure Thermal Shadow for $n=3$ system qubits. The top three qubits in the diagram are the system qubits, and the rest are ancillae. The circuit is delineated by red lines into three sections: (left) random unitary applied to system qubits, (center) QSP operator sequence as generated by pyLIQTR, (right) random Pauli basis transformation on system qubits (ZXX basis in this example).}
    \label{fig:circuit}
\end{figure*}

Based on the background outlined in section \ref{section:bg}, our implementation of Pure Thermal Shadows operates as follows:
\begin{enumerate}
    \item Apply random $n$-qubit unitary $U$ to initial state $\ket{0}^{\otimes n}$
    \item Apply operator $e^{-\beta H/2}$ to the state
    \item Measure the state in a random pauli basis by applying operator $V$ to produce measurement outcome $\ket{b}$
    \item Use $\ket{b}$ to construct a shadow $\hat{\eta}_{V,b}$
    \item Calculate an expectation value estimate for each observable, i.e. $\text{Tr}\hat{\eta}_{V,b}O_j \ \forall O_j \in O$ 
    \item Repeat steps one through five $n_s$ times
    \item Calculate the median-of-means of all expectation value estimates for each observable.
\end{enumerate}
Steps one through three correspond to components of the quantum circuits generated/measured $n_s$ times to construct our shadows (pictured in Figure \ref{fig:PTS}). Note that a circuit like this is necessary for \textit{each shadow} that we produce. We implement our version of these circuits using the Cirq and pyLIQTR packages  \cite{cirq} \cite{pyLIQTR}, an example of the generated quantum circuit is given in Figure \ref{fig:circuit}, which highlights our version the three components described in steps one through three. Now, let us describe in more detail the construction of each component.

\subsection{Random Unitary for TPQ Preparation}
\label{section:RU}
As previously noted, we must sample $U$ uniformly from the $N$-qubit Clifford group such that it forms at least a unitary 2-design. Packages like Qiskit and Stim include random Clifford circuit generators that run in polynomial time/depth (based on ~\cite{cliffSamp}) and form a unitary 3-design. This approach was utilized in the original Pure Thermal Shadow circuit. However, it turns out that more efficient random circuit generators exist if we strictly sample from a unitary 2-design (though any unitary $t$-design for $t \geq 2$ will work). One such algorithm is proposed in ~\cite{dankert} which requires a linear number of gates and logarithmic depth with respect to system size. A summary of this algorithm's procedure is as follows:
\begin{enumerate}
    \item $\mathcal{C}_1/\mathcal{P}_1$-twirl qubit $k$ $\forall k \in \{1, ..., n\}$
    \item Perform random XOR on first qubit
    \item Apply $H$ to the first qubit, and $\mathcal{C}_1/\mathcal{P}_1$-twirl qubit $k$ $\forall k \in \{2, ..., n\}$
    \item Perform random XOR on first qubit
    \item Apply $H$ to the first qubit, and $\mathcal{C}_1/\mathcal{P}_1$-twirl qubit $k$ $\forall k \in \{2, ..., n\}$
    \item Apply $S$ to the first qubit with probability $1/2$
    \item Perform random XOR on first qubit
    \item $\mathcal{C}_1/\mathcal{P}_1$-twirl the first qubit
\end{enumerate}
Note that a $\mathcal{C}_1/\mathcal{P}_1$-twirl consists of randomly applying some gate $R^i$ for $i \in \{0,1,2\}$ to the qubit, where $R = SH$. A random XOR on a qubit means applying several CNOT gates with it as the target, each controlled by another system qubit and applied with probability $3/4$ (or probably $1/4$ that no gate is applied with any given qubit as the control). This operation requires a linear number of gates, however with some clever parallelization, the circuit depth can be reduced to logarithmic scaling with system size. For the sake of efficiency, our implementation utilizes the procedure outlined above in favor of~\cite{cliffSamp}.

\subsection{Non-Unitary operator}
\label{section:NU}
Step 2 requires the application of a non-unitary operator, $e^{-\beta H/2}$ to our quantum state. We can conceptualize this operator as a Quantum Imaginary Time Evolution (QITE) consisting of a Hamiltonian rescaled by its eigenvalues $\tilde{H} = (H - \lambda_{min}I)/(\lambda_{max}-\lambda_{min})$ being evolved for imaginary time $\tau = \beta(\lambda_{max}-\lambda_{min})/2$, like so:
\begin{equation}
e^{-\tau \tilde{H}} = e^{-\beta \lambda_{min}/2} e^{-\beta H/2}
\label{QITE}
\end{equation}
Note that this agrees with our desired non-unitary operator up to a constant factor, which cancels out with normalization when constructing $\ket{\psi _\beta}$ (Equation \ref{tpq}).

In order to (approximately) implement this in a quantum circuit, we utilize Quantum Signal Processing (QSP). QSP allows us to calculate polynomial functions of a Hamiltonian, provided its Block Encoding. Specifically, we calculate a $d$-degree polynomial of the block encoding for $\tilde{H}$ which approximates the exponential function used in equation \ref{QITE}. This polynomial, $\pi_d$ is found by minimizing the $L_{\infty}$-norm of the difference between itself and the function it approximates. Broadly speaking, the approximation is better the higher the degree, but also requires a deeper circuit. The pyLIQTR package includes tools which generate a Block Encoding (via linear Combinations of Unitaries) for $\tilde{H}$, fit our QITE operator to a $d$-degree polynomial $\pi_d(x)$, and produce an operator sequence (with the accompanying phase angles) $\mathcal{O}_\phi$ that takes the Block Encoding of $\tilde{H}$ to the Block Encoding of $\pi_d(\tilde{H})$:
\begin{equation}
\begin{pmatrix} \tilde{H} & \cdot \\ \cdot & \cdot \end{pmatrix} \xmapsto{\mathcal{O}_\phi} \begin{pmatrix} \pi_d(\tilde{H}) & \cdot \\ \cdot & \cdot \end{pmatrix} \approx \begin{pmatrix} e^{-\tau \tilde{H}} & \cdot \\ \cdot & \cdot \end{pmatrix}
\end{equation}
The number of phases (which determines the length of the operator sequence) is $2d + 1$~\cite{pyLIQTR}. As a result, calculating higher degree polynomials will require implementing more operations, resulting in a deeper circuit.

\subsection{Random Measurement Operator}
\label{section:RMO}
Per section \ref{section:CS}, we require random basis measurements to construct shadows. We implement this by applying another random unitary $V$ before the usual computational basis measurement. Random Clifford unitaries are sufficient and can be sampled in polynomial time using the method of~\cite{cliffSamp} (this is also done in the original Pure Thermal Shadow circuit). However, random Pauli basis measurements can be implemented trivially (linear number of gates in constant depth) with minimal performance drawback. Hence, we opt for this approach and $V$ is uniformly sampled from $\{H, HS^{\dagger}, I\}$.

\section{Numerical Simulations}
\label{section:sims}
\begin{figure*}
    \centering
    \includegraphics[width=5.5cm]{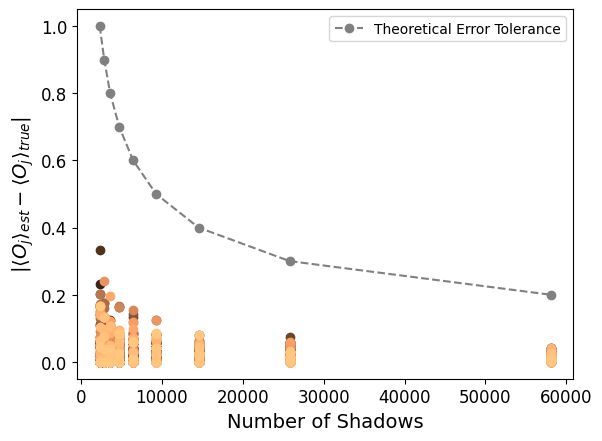}
    \includegraphics[width=5.5cm]{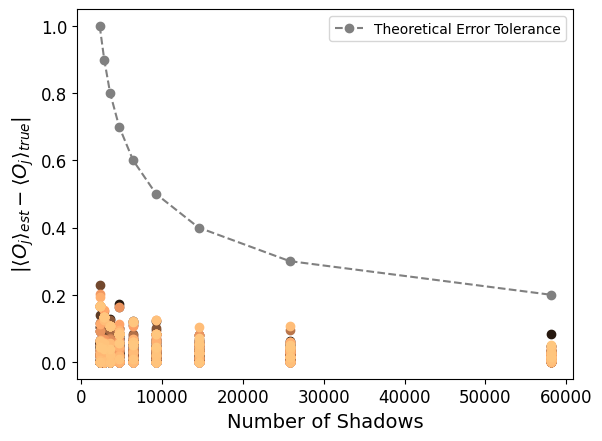}
    \includegraphics[width=5.5cm]{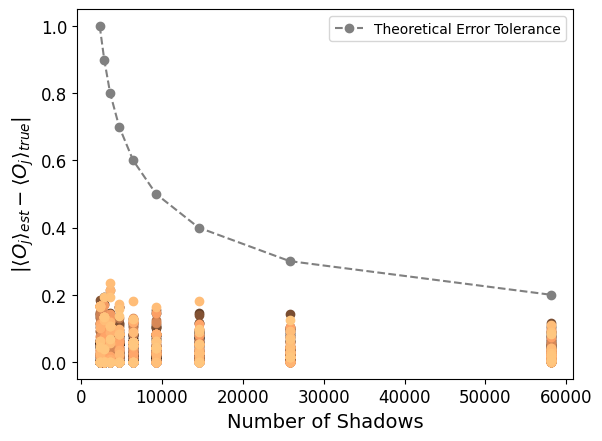}
    \includegraphics[width=7cm]{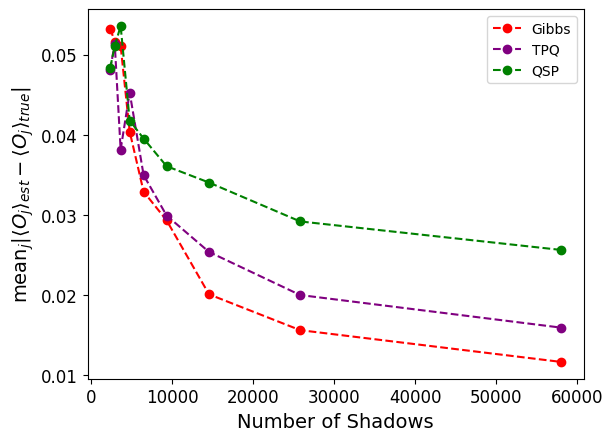}
    \includegraphics[width=7cm]{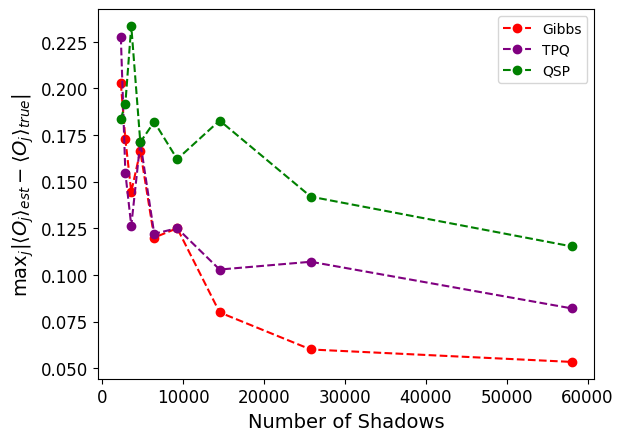}
    \caption{Performance validation for our implementation of Pure Thermal Shadows for a 6-qubit system described by the Hamiltonian in Section \ref{section:sims}.\\ Top: Errors for each observable vs. number of shadows used in estimation for the exact Gibbs state (left) exact TPQ state (center) and TPQ via QSP (right). Note that each color data point corresponds to a different observable (a 1 or 2 qubit Pauli). The lack of any correlation/gradient in the color shows that no particular observable is more prone to errors than the others. Theoretical error tolerance (grey curve) for each number of shadows refers to epsilon in the probabilistic error bound given by equation \ref{numshads} (ignoring the exponentially vanishing term in the mean squared error for TPQ states). In other words, there is a $1 - \delta$ (99\% for these simulations) chance that any given data point will have an error below the grey curve. While providing a worse approximation than shadows of the actual Gibbs state, TPQ states (even approximately prepared through QSP) appear to respect this error bound. \\Bottom: Comparisons of the results from the three plots at the top, including  (left) mean error and (right) maximum error across all observables vs. number of shadows used in the estimations. While errors get worse when using the exact TPQ state and worse still with QSP-prepared TPQ state, they all exhibit the same behavior with respect to the number of shadows used.}
    \label{fig:perfVal}
\end{figure*}

To validate the ability of the approach for approximating Gibbs shadows outlined above, we perform operator-level numerical simulations (not quantum circuits) of both Classical and Pure Thermal Shadows. In particular, we simulate the following: Classical Shadows from measurements of the \textit{exact} Gibbs state, Pure Thermal Shadows from measuring exact TPQ states (i.e. applying the exact non-unitary operation $e^{-\beta H/2}$ to the state before measurement), and finally Pure Thermal Shadows from measuring TPQ states prepared via QSP. For these simulations, we consider the case of an XXZ-Heisenberg Hamiltonian:
\begin{multline}
    H = \sum\limits_{i=1}^{n-1} \left(J_x\sigma^x_{i}\otimes\sigma^x_{i+1}+J_y\sigma^y_{i}\otimes\sigma^y_{i+1}+J_z\sigma^z_{i}\otimes\sigma^z_{i+1}\right) + \\ \sum\limits_{i=1}^{n} \left( h_x \sigma^x_i + h_y \sigma^y_i + h_z \sigma^z_i \right)
    \label{ham}
\end{multline}
for our numerical simulations. We estimate expectation values for Gibbs states of a Hamiltonian as described in equation \ref{ham} at $\beta = 1.5$, with $J_z = 1.0, J_x = J_y = 1.1, h_x = -J_z, h_y = h_z = 0$. Moreover, we choose the set of observables for which we estimate these expectation values, $O$, to be the set of all 1 and 2 qubit Pauli operators (note that as a result, we will consider larger set sizes for increased system sizes). 

The results of these simulations are shown in Figure \ref{fig:perfVal}. All of the simulations show increased performance as the number of shadows used in the estimations increases, and all errors agree with the error bounds described by Equation \ref{numshads}. It also does not appear to be the case that any particular observables are consistently poorly estimated. Moreover, it is generally the case that the "exact" Pure Thermal Shadows perform virtually as well as Classical Shadows (with small deviations owing to the exponentially small bias term in the mean squared error), and slightly worse still when we use QSP. This reflects that TPQs themselves can merely \textit{approximate} Gibbs shadows, and QSP further approximates TPQ state preparation. That said, the Pure Thermal Shadows' deviations from the Classical Gibbs Shadows' estimations are relatively small, and still do not cross the theoretical error bounds---confirming that the approach (even with our deviations from the original) is indeed a valid one for approximating the Classical Shadows of Gibbs states.

\begin{figure}
    \includegraphics[width=8cm]{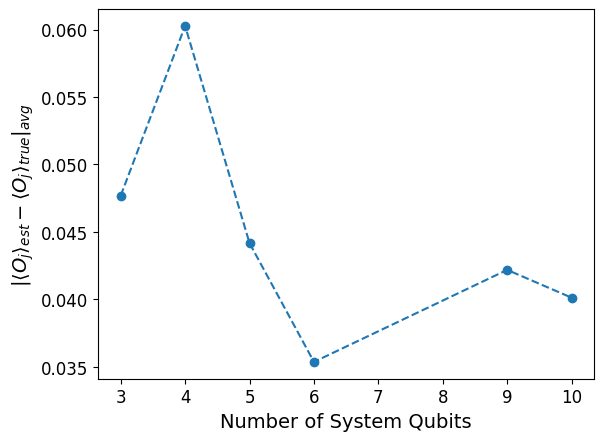}
    \caption{Mean errors for expectation value estimates of 1 and 2 qubit Paulis using Pure Thermal Shadows ($\epsilon = 0.2, \delta = 0.01$) for Gibbs states of varying system sizes. Systems are XXZ Heisenberg Hamiltonians as described by Equation \ref{ham}. Note the lack of a discernible pattern here, and that the maximum error observed, 0.06, falls well within $\epsilon$.}
    \label{fig:PS}
\end{figure}

Now that the performance of Pure Thermal Shadows have been verified, it is also important to consider how the performance of the method scales. Specifically, how can we expect the algorithm to perform as system size is increased? In Figure \ref{fig:PS} we show the average error across expectation value estimates for 1 and 2 qubit Paulis with respect to 3 to 10 qubit Gibbs states for the same Hamiltonian outlined above, using shadows with a fixed $\epsilon = 0.2$. There does not appear to be any meaningful correlation with system size, and all errors fall below $0.2$, obeying the theoretical error bound. This suggests that as we subsequently consider the resources required to construct Pure Thermal Shadows for various system sizes, we should not necessarily expect to find (predictably) different performances associated with them.

\section{Resource Analysis}
\label{section:RA}
\begin{figure}
    \centering
    \includegraphics[width=8cm]{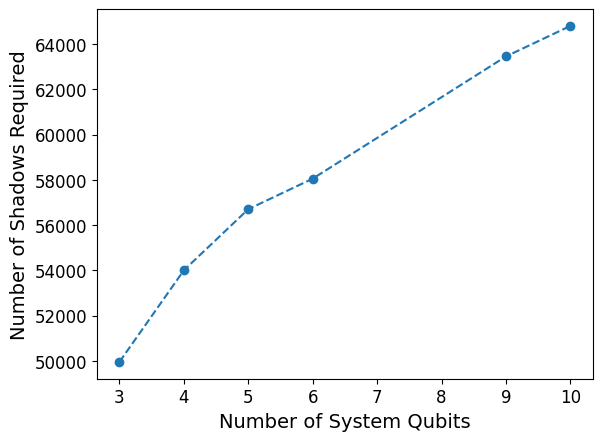}
    \includegraphics[width=8cm]{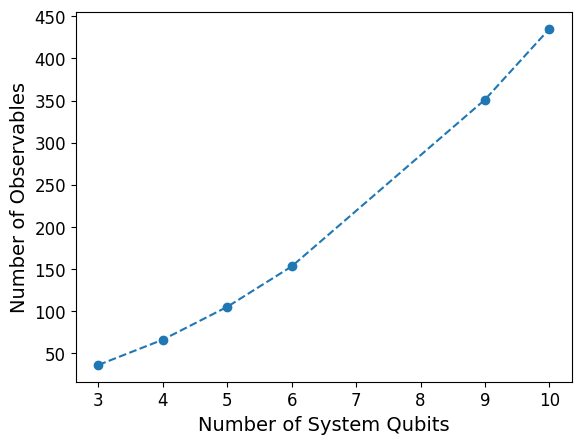}
    \includegraphics[width=8cm]{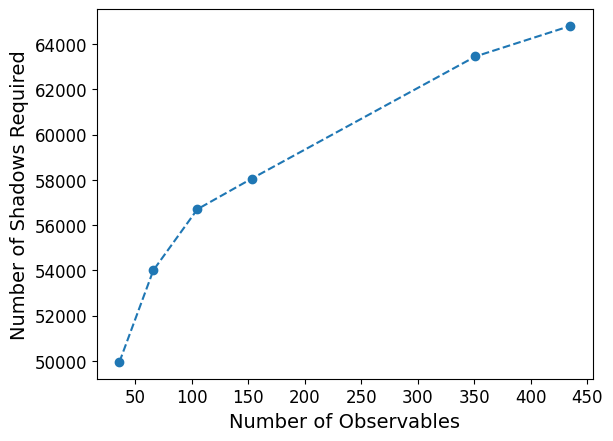}
    \caption{Top: The number of shadows required for $\epsilon$ = 0.2, $\delta=0.01$ for different system sizes. \\ Center: How the number of possible 1 and 2-qubit Pauli operators (our observable set) scales with system size, as described in Section \ref{section:RA}. \\ Bottom: Per Equation \ref{numshads}, the number of shadows necessary (for the given $\epsilon$, $\delta$) as a function of the size of our set of observables. Note that the number of observables for each data point corresponds to a system size in the other two plots (i.e. first point corresponds to 3-qubit system, second 4, et cetera). Also note that the slight deviation from the trend at 6 qubits (153 observables) is due to rounding $S$ in $n_s = SK$ to the nearest integer, so that the shadows can be properly partitioned into sets for median of means estimation.}
    \label{fig:SSC}
\end{figure}

There are a number of factors that determine the difficulty of estimating Gibbs state properties with our implementation. The resources associated with each quantum circuit depend on the size of the system as well as the degree of polynomial used in our QSP approximation. Furthermore, the number of unique circuits we must construct and measure depends on the desired accuracy (as described in Section \ref{section:PTS}). Recall the result that for a set of $M$ observables we only require $O(log(M))$ shadows for estimation. Even while restricting our set of observables to one and two qubit Paulis, as system size increases there are more such possible operators because there are more places to 'put' the non-identity Paulis. Specifically: $M = 3n + 3^2 {n \choose 2}$ where the first term is the number of one qubit Paulis and the second the number of two qubit Paulis. While the number of shadows (and thus number of circuits to measure and generate) does not explicitly depend on system size, plugging this expression into Equation \ref{numshads} provides a sense for how many shadows are necessary for specifically estimating one and two qubit Paulis of an $n$-qubit Gibbs state as a function of $n$. As such, an analysis of the required number of measurements as functions of both the number of system qubits and number observables is shown in Figure \ref{fig:SSC}. The aforementioned logarithmic dependence on $M$ is clear, however we can see that the total number of shadows required is large, on the order of tens of thousands. This is largely a consequence of the inverse square dependency on $\epsilon$ per Equation \ref{numshads} compared with the logarithmic dependence on $M$. To illustrate how grave this effect is, consider a case of relatively small $M$; let us say we aim to estimate two observables each with locality of two, but with $\epsilon = 0.2$ and $\delta = 0.01$. According to Equation \ref{numshads}, this would still require just over 32,000 shadows. This suggests that for a small number of observables, a shadow tomography-based approach might not be measurement efficient, the only benefit being its independence from system size. Thus, to see the benefits in measurement complexity with Pure Thermal Shadows would require looking at large systems and/or many observables.

\begin{figure*}
    \centering
    \includegraphics[width=5.7cm]{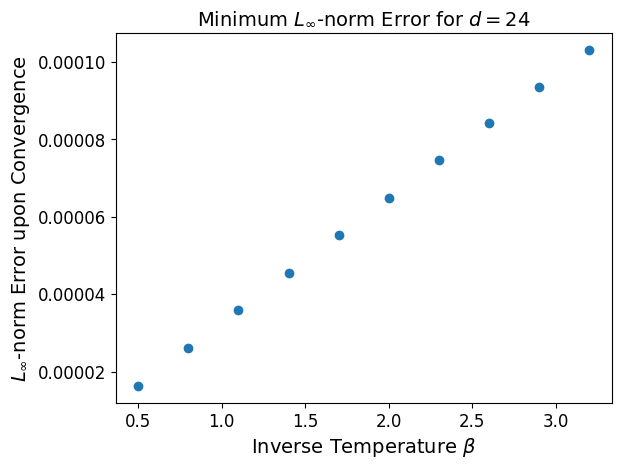}
    \includegraphics[width=5.7cm]{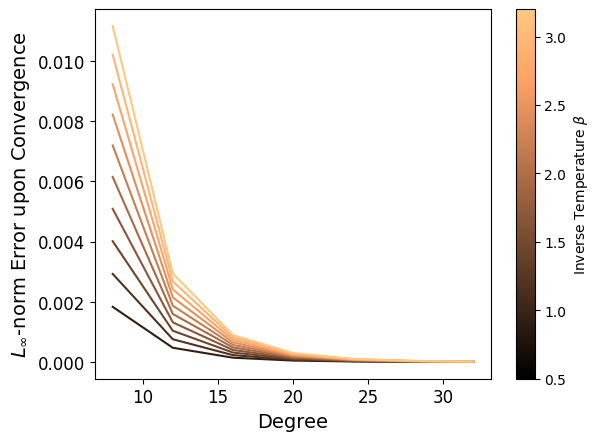}
    \includegraphics[width=5.7cm]{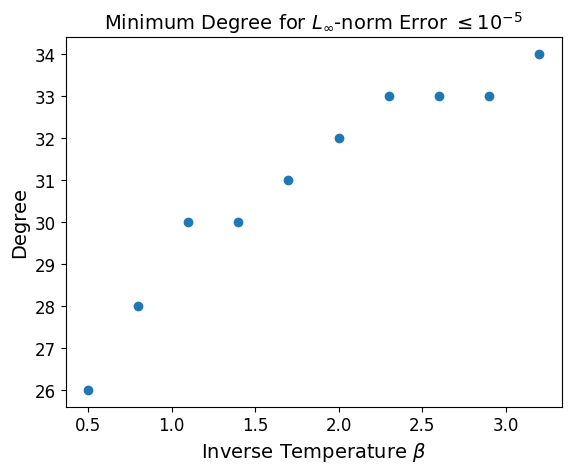}
    \caption{Results for the approximation quality of $\pi_d(x)$ with respect to $e^{-\beta x}$, the general form of our QITE operator, agnostic to specific Hamiltonians. \\ Left: The $L_{\infty}$-norm error of the best polynomial $\pi_{24}(x)$ approximating the function $e^{-\beta x}$ for increasing $\beta$; the error increases linearly with $\beta$ \\
    Center: Error for best $\pi_d$ with increasing $d$ and $\beta$. Error converges with respect to $d$ more slowly for increasing $\beta$; in other words, increasing $\beta$ requires a higher degree polynomial to achieve the same error. \\
    Right: The minimum degree $d$ necessary to approximate $e^{-\beta x}$ with an $L_{\infty}$-norm error threshold of $10^{-5}$. While some $\pi_d$ are good enough for multiple $\beta$ (e.g. $d=34$ fits the threshold for $\beta = 2.3, 2.6, 2.9$), overall the degree necessary increases linearly with $\beta$.}
    \label{fig:BDD}
\end{figure*}

As discussed in Section \ref{section:NU}, the degree of the approximating polynomial $\pi_d$ determines the number of operators in a QSP operator sequence, and larger $d$ provides better approximation at the expense of a longer sequence. Naturally, we should consider how the features of the function we are approximating determine the difficulty to perform the approximation. In the case of our QITE operator, we are calculating a function of our rescaled Hamiltonian. As a result, the only part of the function itself which comes from parameters of the Gibbs state is the imaginary time $\tau$ coming from the inverse temperature $\beta$. Let us consider how $\beta$ affects the quality of $\pi_d$'s approximations. Recall that $\pi_d$ is defined as having the smallest possible $L_\infty$-norm difference from the true function. If we consider a $\pi_d$ with a fixed $d$, then the $L_\infty$-norm difference will increase linearly with $\beta$, as depicted in Figure \ref{fig:BDD} for the case $d = 24$. Additionally, we know that as $d$ increases, this error should decrease; this can be observed in combination with the $\beta$-dependence in Figure \ref{fig:BDD}. For higher values of $\beta$, the slower convergence of the $L_\infty$-norm to zero indicates that a higher degree polynomial is required. Figure 2 in~\cite{Coopmans} supports this as well, showing a similar correlation for the actual performance of shadows with respect to $\beta$ and $d$.

Broadly speaking, the relationship between $d$ and $\beta$ for a given $L_\infty$ error threshold is linear (e.g. Figure \ref{fig:BDD} for $||e^{-\beta x} - \pi_d(x)||_\infty \leq 10^{-5}$). Consequently, we can say that since the length of the QSP operator sequence increases linearly with $d$, it also increases linearly with $\beta$--suggesting that the depth of the QSP circuit will increase at least linearly with $\beta$. This is unsurprising considering that the imaginary "time" for which we are evolving our system via QSP is directly proportional to $\beta$, and it is known that the complexity of QSP scales linearly with time~\cite{low}.

With the considerations for sampling complexity and difficulty of approximating QITE in mind, let us more concretely gauge the resource requirements for the circuits used for generating Pure Thermal Shadows. For this purpose, we use pyLIQTR to generate (but not necessarily simulate) quantum circuits in Cirq and decompose them for resource estimation. While pyLIQTR focuses on logical resource estimations (i.e. not considering physical costs associated with error correction procedures), we can analyze the generated circuits to provide resource estimates for  both Noisy Intermediate-Scale Quantum (NISQ) and Fault Tolerant devices. For a Fault Tolerant device, a quantum circuit would be implemented with a Clifford+T gate set, with the T gate being the most resource-intense gate. For a NISQ device, we use single-qubit rotation gates and two-qubit gates, with the latter being more resource-intense. Thus, we decompose Pure Thermal Shadow circuits into Clifford+T (FT) and rotations + 2-qubit gates (NISQ) to get logical resource estimates in the form of gate counts and circuit depths.

\begin{figure*}
    \centering
    \includegraphics[width=7cm]{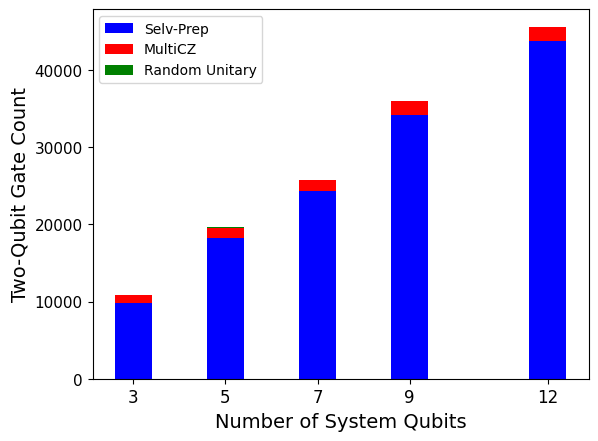}
    \includegraphics[width=7cm]{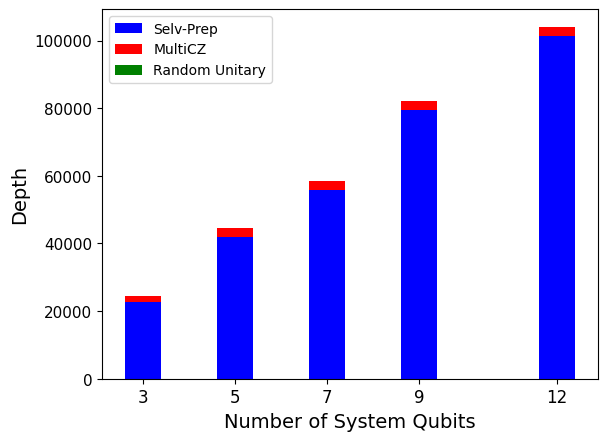}
    \includegraphics[width=7cm]{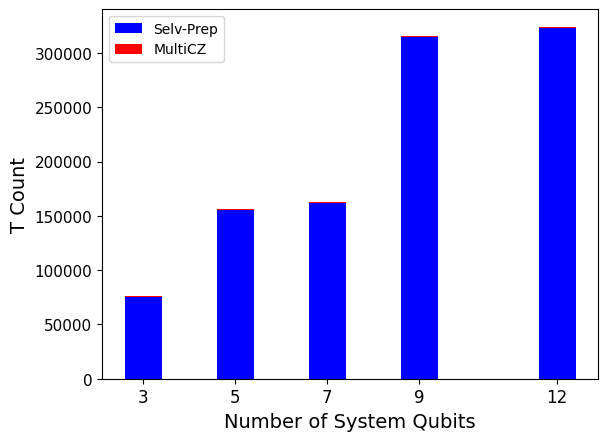}
    \includegraphics[width=7cm]{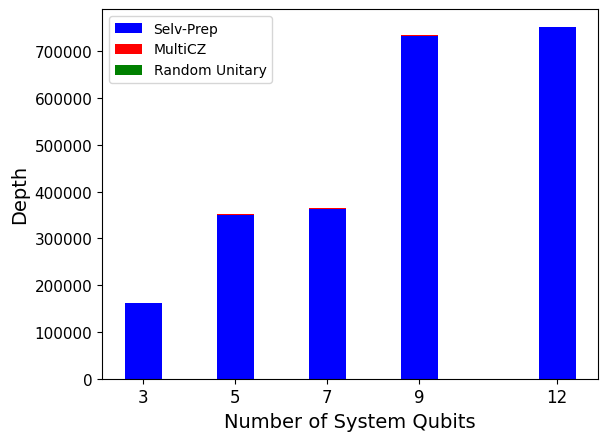}
    \caption{Resource scaling for a circuit used to generate a single shadow of a Gibbs state with the Hamiltonian and parameters described in section \ref{section:sims}. Top: Results for NISQ-like decomposition. For both Two-qubit gate count (left) and total circuit depth (right), QSP operations (Select blocks, and Multi-CZ gates) take up the majority of resources and scale linearly with system size. Random Unitaries take up a comparatively negligible amount of resources (barely visible in the bar plot as they are on the order of at most hundreds) and scale logarithmically with system size (see Appendix A for more detailed results on the Random Unitary). \\
    Bottom: Results for full Clifford+T decomposition (as would be implemented on a Fault Tolerant device). Both T Count and total circuit depth (left and right respectively) exhibit similar behavior to the NISQ decomposition, but have counts an order of magnitude greater ($10^5$ vs. $10^4$).}
    \label{fig:RA}
\end{figure*}

Now, recall that the circuits in question that we aim to decompose consist of three main sub-circuits: 
\begin{enumerate}
    \item Random unitary circuit
    \item QSP circuit implemented in pyLIQTR
    \item Random Pauli basis measurement 
\end{enumerate}
The last component can be easily implemented in constant depth / linear number of gates (at most two gates per qubit; see section \ref{section:RMO}), relatively negligible and omitted from our resource analysis. The QSP circuit is composed of SelectV-Prep blocks and Multi-CZ gates. We perform decompositions for Pure Thermal Shadow circuits of varying system sizes and report gate count / depths for the Select, Multi-CZ, and random unitary components of our circuits. Estimates for Gibbs states of a system described in section \ref{section:sims} utilizing a 24 degree polynomial, in the QSP circuit, are given in Figure \ref{fig:RA}. These results indicate that the QSP components require the most resources for both Fault Tolerant and NISQ estimates, and scale linearly with system size. Overall, we find that in the range of 3-12 qubits, a NISQ device would require on the order of $10^4$ gates/circuit depth for a single shadow and a Fault Tolerant device would require on the order of $10^5$ logical gates/circuit depth. To estimate a desired set of expectation values, we will need to repeat these single circuit executions a number of times. The results shown in Figure~\ref{fig:SSC} give an indication as to the number of repeated measurements required.
Also recall that for greater accuracy and/or to describe a lower-temperature system, we will require even greater resources owing to the necessity of a higher degree polynomial. Nevertheless, these results give some idea of how the necessary resources required per shadow scales with system size.

\section{Discussion}

The results of our resource analysis have major ramifications when considering the utility of Pure Thermal Shadows. First, these results indicate resource necessities which far exceed the capabilities of NISQ devices. As depicted in Figure \ref{fig:RA}, a single Pure Thermal Shadow circuit would require on the order of $10^4$ two-qubit gates and circuit depth, potentially even more for higher values of $\beta$. Even with a high-performance two-qubit gate fidelity like 99.5\%~\cite{eve}, performing this many computations, would result in a very small likelihood of an accurate computation: ($0.995^{10^4} \approx 0$). 

Now, since the scheme for generating Pure Thermal Shadows is certainly not a NISQ algorithm, how does it compare to classical and fault tolerant ones? The dependence on $\beta$ suggests that the algorithm will perform most efficiently for higher temperature systems, for which polynomial time classical Gibbs sampling algorithms already exist. More specifically,~\cite{yin} suggests that an exponential quantum speedup would only be possible for low-temperature systems. Bearing this in mind, it is important to note that the complexity of the fault tolerant algorithm proposed in~\cite{chowdhury} scales linearly with $\sqrt{\beta}$, while our implementation seems to scale linearly only with $\beta$. On the other hand, the former's complexity also scales with the square root of Hilbert space dimension, while our implementation scales with the number of system qubits, meaning it exhibits better scaling with system size. Thus, this algorithm's utility for fault tolerant devices would depend on both temperature and system size; for lower temperature systems of a fixed system size, other algorithms would likely be more efficient, but if the system size is large enough for a given temperature, we could find the opposite.

\section{Conclusion}
In this work, we proposed a modified implementation of Pure Thermal Shadows which was utilized for resource analysis. Our implementation choices that differed from the original algorithm included using random Pauli basis measurements as opposed to random Clifford measurements of the Thermal Pure Quantum states, as well as preparing said states using a more efficient random unitary generator which formed a unitary 2-design as opposed to 3-design. Our approach confirmed the result of the original Pure Thermal Shadow experiments (evidenced via numerical simulations), and these decisions ultimately resulted in a smaller number of operations and circuit depth, providing a "best case scenario" for resource estimates. Even still, our analysis of these resource estimates found large resource requirements for quantum circuits that far exceeded near term capabilities even for small systems, while also scaling linearly with system size. Furthermore, the best estimates will require on the order of $10^4$ measurements of these costly circuits.

While our results preclude the algorithm's utility in the NISQ-era, the possibility of usefulness for a Fault Tolerant device remains---with some constraints. Efficient classical Gibbs sampling algorithms exist for high temperature systems, so this algorithm would only have potential in a low temperature regime. Furthermore, other Fault Tolerant quantum algorithms exist which run in $\sqrt{N \beta}$ time (where $N=2^n$, Hilbert space dimension), featuring a more favorable dependence on temperature but worse dependence on system size than the one found here. Thus, the potential utility of this Pure Thermal Shadows implementation is further constrained to large, low-temperature systems. Fortunately, our results indicate that performance should scale favorably to larger system sizes. 

We must also emphasize the observation that QSP demands the most resources of any part of this algorithm. In contrast, the random unitary generation contains a small number of relatively basic operations whose circuit depth scales logarithmically with system size, and the random Pauli basis measurement operations require only a constant circuit depth of 2. While alternative approaches to QSP exist, such as asymptotically more efficient Block Encodings than the LCU method, these would not provide any benefits for the smaller systems tested here---meaning the utility constraint to larger system sizes would still hold.

Of course, the classical-shadow-inspired sampling procedure here is clearly a useful and efficient one \textit{provided} an (approximately) Gibbs-like state from which to sample. Thus, an interesting question to consider is how we could accomplish this with a less costly alternative to Quantum Signal Processing? Perhaps Pure Thermal Shadows could be generated with a procedure other than Quantum Imaginary Time Evolution. Even turning back to the possibility of NISQ utility, perhaps the random unitary procedure could serve as a useful initilization technique to a variational approach combined with classical shadows. Ultimately, while we have heavily constrained the utility of our particular implementation, more opportunities remain for similar sampling techniques utilizing Pure Thermal Shadows.

\bibliographystyle{abbrv}
\bibliography{refs}

\section*{Acknowledgment}
The authors would like to thank Justin Elenewski for his insights and support on this work. We would also like to thank Josh Mutus and Mark Hodson for introducing us to Pure Thermal Shadows.

\section*{Appendix A - Detailed Resource Estimation for Random Unitary Circuit}
\begin{figure}
    \centering
    \includegraphics[width=8cm]{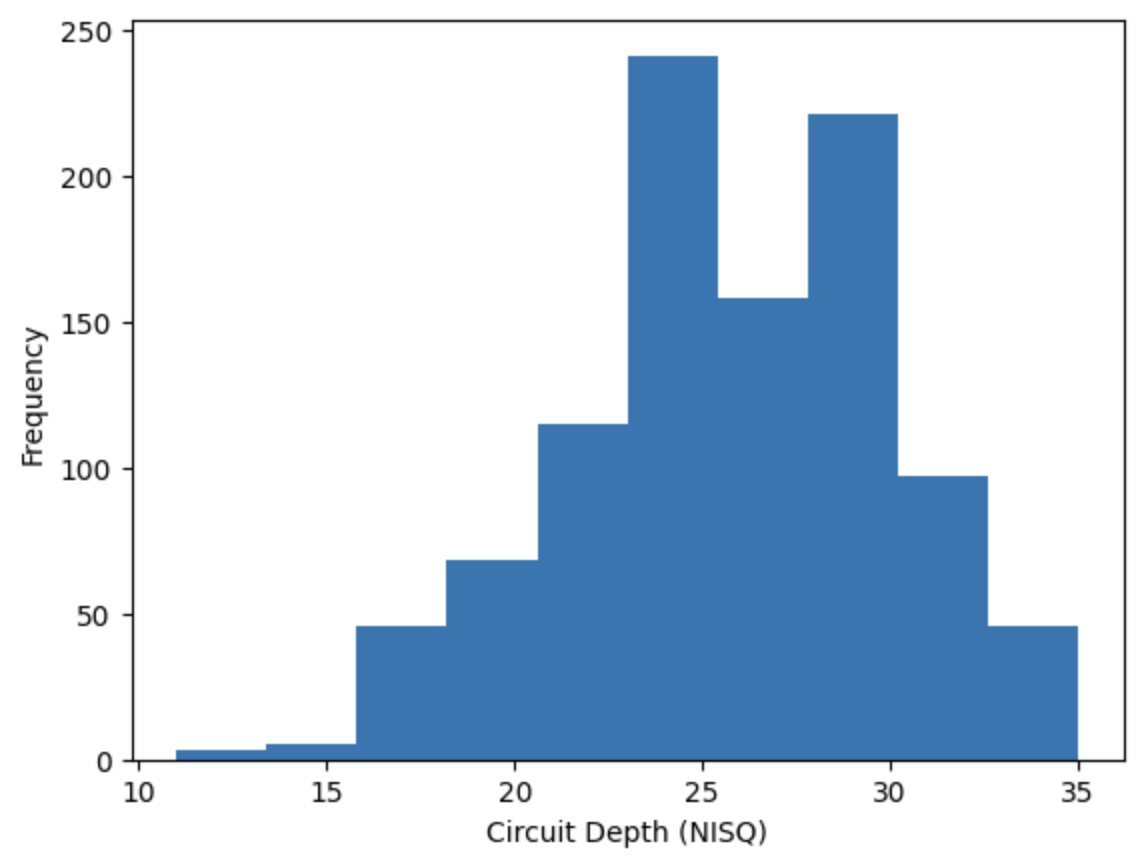}
    \includegraphics[width=8cm]{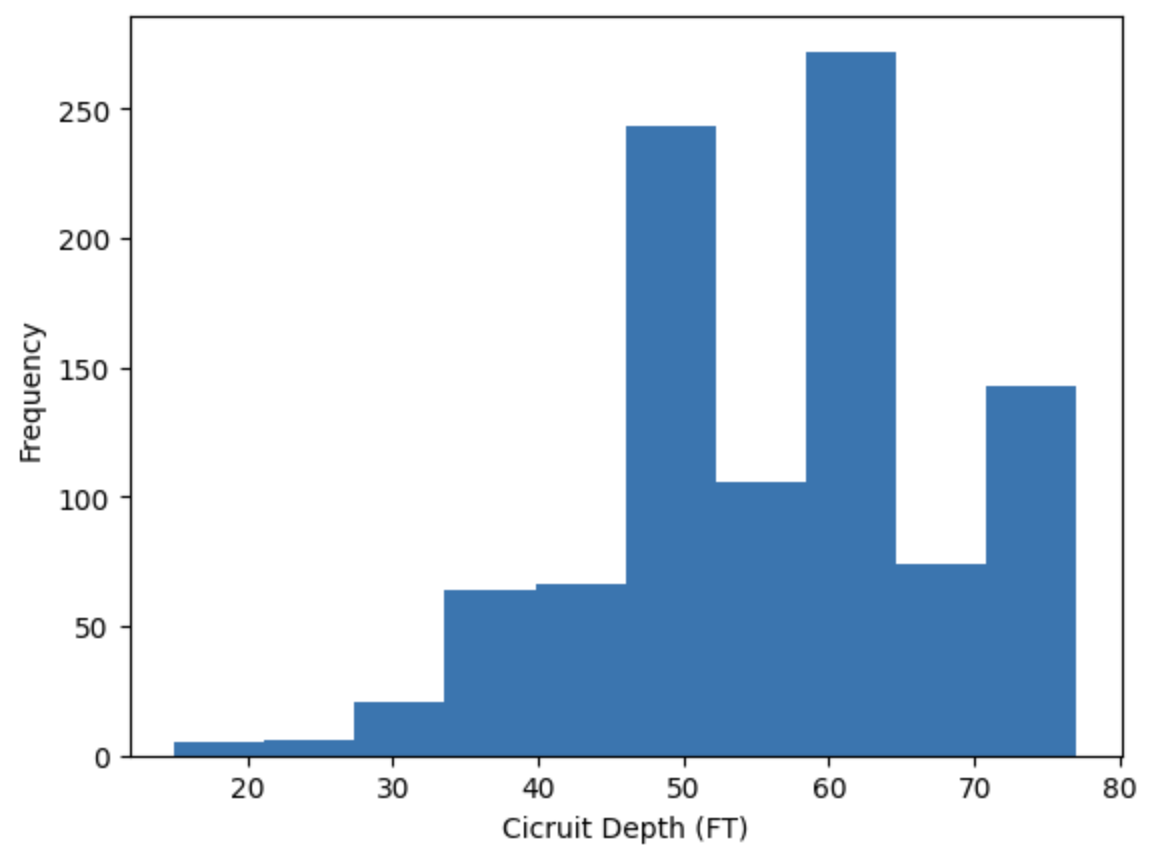}
    \caption{Distribution of random 3-qubit unitary circuit depths based on 1000 samples for (top) NISQ decomposition having a mean of about 25.7 and a standard deviation of about 4.3, and (bottom) Fault Tolerant decomposition having a mean of about 56.1 and standard deviation of about 11.7.}
    \label{fig:ruhist}
\end{figure}

\begin{figure}
    \centering
    \includegraphics[width=8cm]{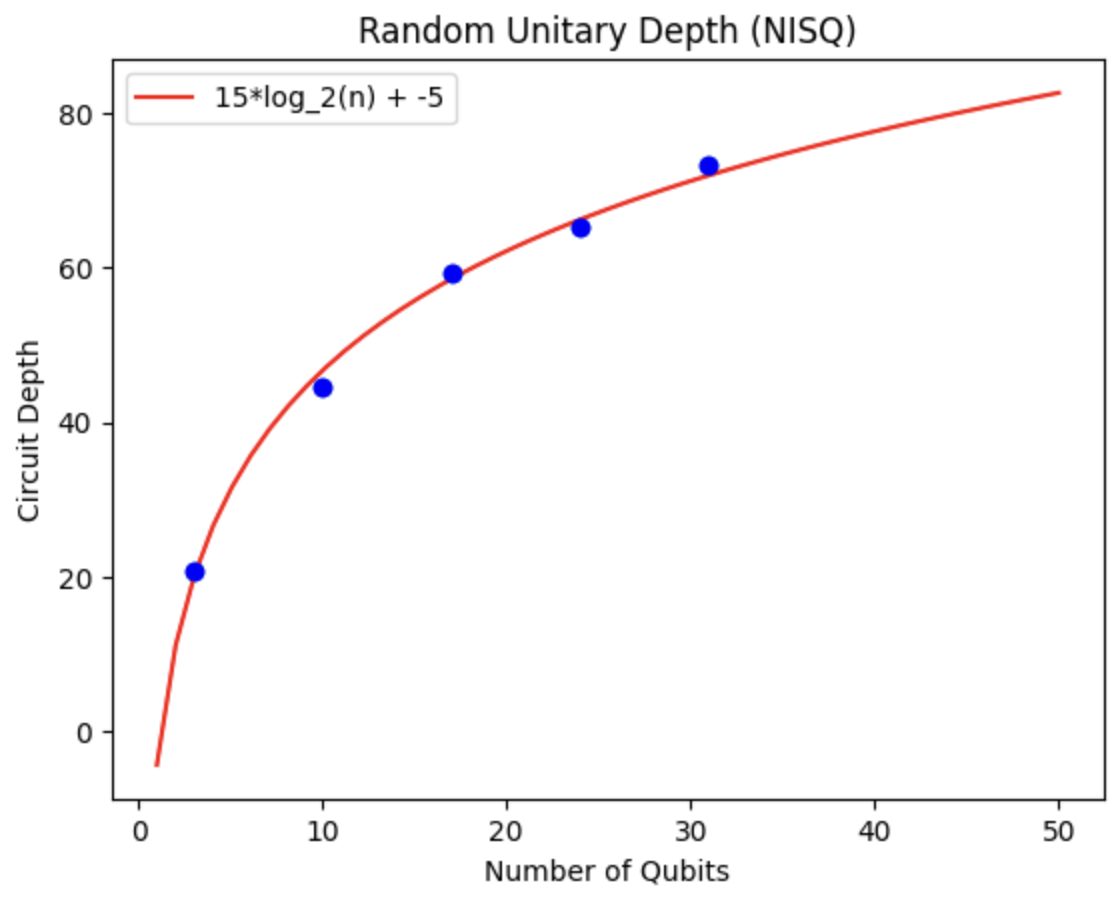}
    \includegraphics[width=8cm]{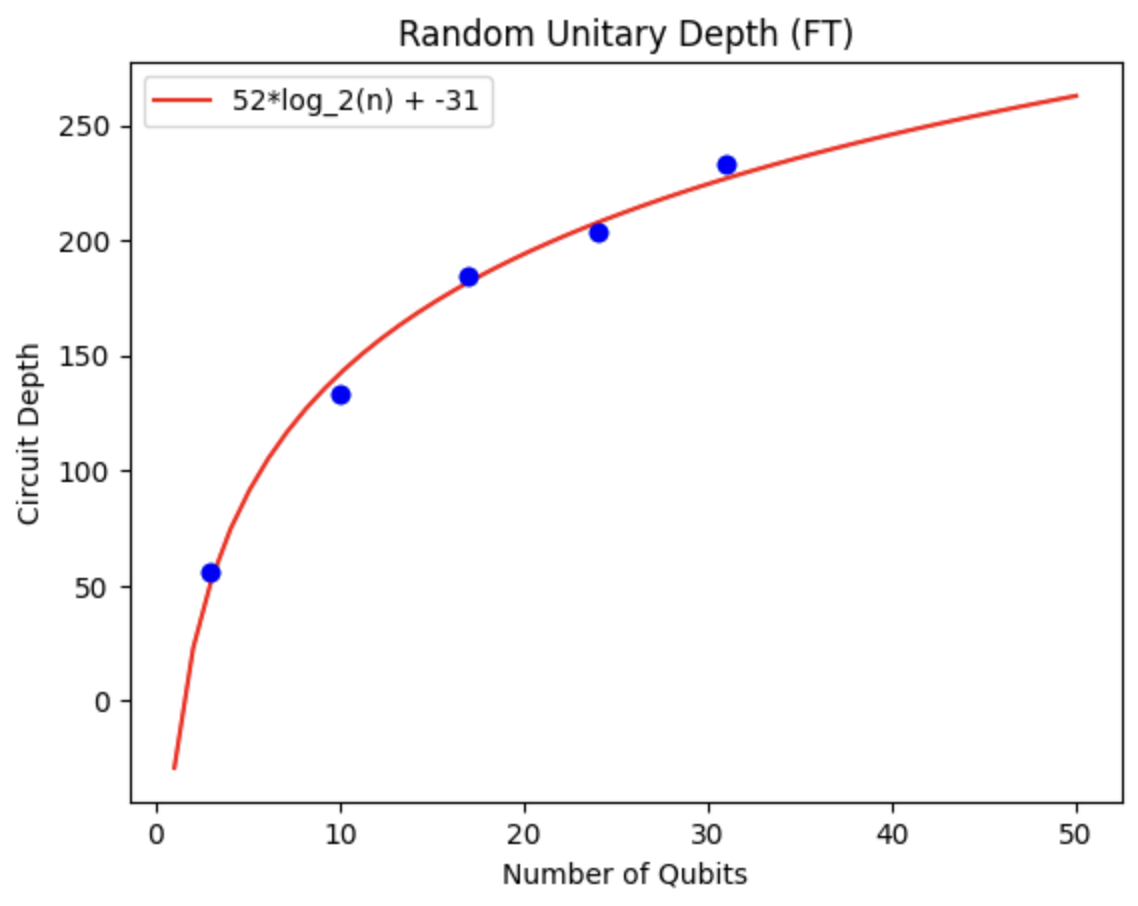}
    \caption{Mean depth of 1000 sampled random unitary circuits for various system sizes for NISQ (top) and Fault Tolerant (bottom) decompositions. The results exhibit the logarithmic scaling behavior as expected, and some lines of fit have been included to highlight this.}
    \label{fig:ruscale}
\end{figure}

In Section \ref{section:RA}, we did not elaborate on exact resource estimation results for the random unitary circuit (procedure adapted from~\cite{dankert}), and they were not apparent from our figures as they paled in comparison to the QSP resources. For completion, we report some more detailed results here. As depicted in Figure \ref{fig:ruhist}, we see that for both NISQ and Fault Tolerant decompositions, the depth of the random unitary circuit follows a relatively normal distribution with some negative skew. When plotting mean circuit depths for varied system sizes in Figure \ref{fig:ruscale}, we can also observe the logarithmic relationship described in Section \ref{section:RU}.

\end{document}